\documentstyle[12pt]{article}

\renewcommand{\thefootnote}{\fnsymbol{footnote}}

\newcommand{\be}{\begin{equation}}
\newcommand{\ee}{\end{equation}}

\newcommand{\bea}{\begin{eqnarray}}
\newcommand{\eea}{\end{eqnarray}}

\newcommand{\NP}[1]{Nucl. Phys.\ {\bf #1}\ }
\newcommand{\PL}[1]{Phys. Lett.\ {\bf #1}\ }

\newcommand{\PR}[1]{Phys. Rev.\ {\bf #1}\ }
\newcommand{\PRL}[1]{Phys. Rev. Lett.\ {\bf #1}\ }

\newcommand{\MPL}[1]{Mod. Phys. Lett.\ {\bf #1}\ }

\def\lsim{\raise0.3ex\hbox{$<$\kern-0.75em\raise-1.1ex\hbox{$\sim$}}}
\def\gsim{\raise0.3ex\hbox{$>$\kern-0.75em\raise-1.1ex\hbox{$\sim$}}}

\newcommand{\B}{\beta}
\newcommand{\lm}{\lambda}

\begin{document}
\begin{titlepage}
\begin{flushright}
{\large \bf HU-SEFT R 1995-12}
\end{flushright}
\vskip 2cm
\begin{center}
\vskip .2in

{\Large \bf
On Quadratic Divergences and the Higgs Mass}
\vskip .4in

Masud Chaichian$^{a,b}$\footnote{
E-mail: chaichian@phcu.helsinki.fi},
Ricardo Gonzalez Felipe$^{b}$\footnote{
E-mail: gonzalez@phcu.helsinki.fi} and
Katri Huitu$^{b}$\footnote{
E-mail: huitu@phcu.helsinki.fi}\\[.15in]

{\em $^{a}$Laboratory for High Energy Physics,
     Department of Physics}

{\em $^{b}$Research Institute for High Energy Physics}\\[.15in]

{\em
P.O. Box 9 (Siltavuorenpenger 20 C)\\
FIN-00014 University of Helsinki, Finland}

\end{center}

\vskip 1.5in

\begin{abstract}

The quadratic divergences in the scalar sector of the standard
model are considered.
Since the divergences are present also in the unbroken theory,
a natural scale for the divergence formula is proposed to be at
the scale of new physics.
The implications of top quark mass on the Higgs mass are
investigated by means of the renormalization group equations. The
Coleman-Weinberg mechanism for spontaneous symmetry breaking is
also considered.
\end{abstract}
\end{titlepage}

\newpage
\renewcommand{\thepage}{\arabic{page}}
\setcounter{page}{1}
\setcounter{footnote}{1}


Supersymmetric theories have attracted a great deal of attention mainly
due to the fact that they have the property of being free of quadratic
divergences. It is therefore natural to ask whether the cancellation of
such divergences may occur in nonsupersymmetric theories as well and,
in particular, in the standard model.

According to the discussion given by Veltman \cite{mV},
suggestive of an underlying theory with a symmetry
protecting the mass, the quadratic
divergences in the standard model should cancel as happens with the
electron and gauge boson masses.
For the masses in the broken phase this implies the relation \cite{mV}
$$
\frac{3}{2}m_W^2+\frac{3}{4}m_Z^2+\frac{3}{4}m_H^2=\sum_f m_f^2 ,
$$
where $f$ stands for the fermions. Indeed this
relation follows from the relation among the coupling constants:

\be
\label{eqvelt}
\frac 32 g_1^2 +\frac 92 g_2^2
+ 6 \lambda = 4 \sum_f h_f^2 \ ,
\ee

\vspace{.2truecm}
\noindent
with $g_1$ and $g_2$ being the $U(1)$ and $SU(2)$
gauge couplings respectively, $\lambda $  the
Higgs self-coupling and $h_f$ the Yukawa couplings.

Much work on the mass formulas has been done in details
and in different aspects in
refs.~\cite{JJ}-\cite{CP}.
In \cite{RGV,CMM}, using the dimensional regularization the quadratic
divergences have been put to zero in two-loop order to
determine
unambiguously the top and Higgs masses.
In this case, however, the equations from the
higher loop corrections are not compatible with the formulas from
the lower loop orders \cite{JJ}.
Equating, in addition to quadratic divergences, one of the logarithmic
divergences to zero has also been used to find a second
equation \cite{DP,OW,CP}.
For this purpose, logarithmic divergences of either the Higgs self-energy,
or electron self-energy, or $eeH$ coupling have been considered. For
instance,
in the papers by Osland and Wu \cite{OW}, by imposing the cancellation of
quadratic divergences and the logarithmic divergences in the $eeH$ vertex,
 the Higgs and top masses were
determined to be $m_H\sim 190$~GeV and $m_t\sim 120$~GeV.

The physical motivation of these approaches seems not to be quite
satisfactory:
one problem with the mass formulas found from the
vanishing of the divergences is that the formulas are
not invariant under the renormalization group transformations.
Thus the formula (\ref{eqvelt}) is defined at some scale, which is not
determined.
In \cite{AJJ}, it was shown that the formula (\ref{eqvelt})
can be required to be scale independent provided that strong interactions
are ignored.
Taking the strong interactions (which give important contribution) into
account, one does not find a scale independent solution.

In this letter we study the formula (\ref{eqvelt}) as a
function of the scale. We shall not consider the
contributions from leptons and lighter quarks which are negligible
in Eq.(1).
Since the quadratic divergences exist already before the symmetry
breaking, it is natural to require that the equation should be valid
at a large scale $\Lambda$.
The scale is not a priori determined, since we do
not know when new physics enters into play.
The corrections due to (\ref{eqvelt}) should not however
exceed the physical scalar mass.
We consider the consequences of these requirements
on the predictions for the Higgs mass at the electroweak scale.
Furthermore, we shall consider the possibility of combining the
Coleman-Weinberg idea \cite{CW} for the spontaneous symmetry breaking
with Veltman's idea of cancellation of quadratic divergences in the
standard model.

\section{RGE and Quadratic Divergences}

Since Eq. (\ref{eqvelt}) is not invariant under the
renormalization group transformations,
one should take into account the running of the couplings according to
the renormalization group equations (RGE) in
finding the physical masses at the electroweak
scale.

The RGE up to  two loops for the gauge couplings \mbox{$g_i, i=1,2,3$}
($g_3$ is the strong coupling),
the top Yukawa coupling $h_t$
and the scalar coupling $\lm $ are given by \cite{GG}-\cite{FJS}

\be
\label{eqsuthree}
{\frac {dg_i}{dt}}= {\frac 1{16\pi ^2}} \B_i   \;\;\;,\;
{\frac {dh_t}{dt}}= {\frac 1{16\pi ^2}}\B_t \;\;\;,\;
{\frac {d\lm }{dt}}= {\frac 1{16\pi ^2}}\B_\lm \;,
\ee

\vspace{.3truecm}
\noindent
where $t=\ln( \mu /\mu_0 )$ and

\bea
\B_1 &=& \frac{41}{6} g_1^3 + \frac{1}{16\pi^2} \left\{ \frac{199}{18}g_1^2
 + \frac{9}{2}g_2^2+\frac{44}{3}g_3^2-\frac{17}{6}h_t^2 \right\} g_1^3 \ ,
\nonumber\\
\B_2 &=& -\frac{19}{6} g_2^3 + \frac{1}{16\pi^2} \left\{ \frac{3}{2}g_1^2
 + \frac{35}{6}g_2^2+12g_3^2-\frac{3}{2}h_t^2 \right\} g_2^3 \ ,
\nonumber\\
\B_3 &=& -7g_3^3 +
\frac{1}{16\pi^2} \left\{ \frac{11}{6}g_1^2+\frac{9}{2}g_2^2-26g_3^2-2h_t^2
\right\} g_3^3 \ ,\nonumber \\
\B_t &=&
\left({\frac 92}h_t^2 -{\frac{17}{12}}g_1^2
-{\frac 94}g_2^2 -8 g_3^2 \right) h_t \nonumber \\
     & & + \frac{1}{16\pi^2} \left\{
-12h_t^4 + h_t^2 \left( \frac{131}{16}g_1^2 +\frac{225}{16}g_2^2
+ 36g_3^2 -6\lm \right) + \frac{3}{2}\lm^2 \right. \nonumber\\
     & & \left. +\frac{1187}{216}g_1^4 -\frac{3}{4}g_1^2g_2^2
+\frac{19}{9}g_1^2g_3^2
-\frac{23}{4}g_2^4+9g_2^2g_3^2-108g_3^4 \right\}h_t \ , \nonumber \\
\B_\lm &=&
 12\lm ^2- ( 3 g_1^2 +9 g_2^2
-12h_t^2)\lm +
{\frac 34}g_1^4 +{\frac{3}{2}}g_1^2g_2^2
+{\frac 94}g_2^4 -12 h_t^4 \nonumber\\
     & & + \frac{1}{16\pi^2} \left\{ -78\lm^3 -72\lm^2h_t^2 +18\lm^2(g_1^2+
3g_2^2)-3\lm h_t^4+80\lm g_3^2h_t^2 \right. \nonumber \\
     & & +\frac{45}{2}\lm g_2^2h_t^2+\frac{85}{6}\lm g_1^2h_t^2
-\frac{73}{8}\lm g_2^4+\frac{39}{4}\lm g_1^2g_2^2+\frac{629}{24}\lm g_1^4
\nonumber \\
     & & +60h_t^6-64h_t^4g_3^2 - \frac{16}{3}h_t^4g_1^2
-\frac{9}{2}h_t^2g_2^4+21h_t^2g_1^2g_2^2 \nonumber \\
     & & \left. -\frac{19}{2}h_t^2g_1^4+\frac{305}{8}g_2^6
-\frac{289}{24}g_1^2g_2^4
-\frac{559}{24}g_1^4g_2^2-\frac{379}{24}g_1^6 \right\} .
\eea

\vspace{.3truecm}
\noindent

Neglecting the two-loop contributions, which from our further analysis
 turn out to be of the order of a few percent, the one-loop equations
(\ref{eqsuthree}) for gauge
couplings can be solved analytically and their
solutions read as

\bea
g_1^2 (\mu )& =& {\frac {g_1^2 (\mu _0)}
{1-{\frac {41}{48\pi^2}} g_1^2(\mu_0 )\, {\rm ln }(\mu /\mu_0 )}}\ ,
\nonumber\\
g_2^2 (\mu )& =& {\frac {g_2^2 (\mu _0)}
{1+{\frac {19}{48\pi^2}} g_2^2(\mu_0 )\, {\rm ln }(\mu /\mu_0 )}}\ ,
\nonumber\\
g_3^2 (\mu )& =& {\frac {g_3^2 (\mu _0)}
{1+{\frac {7}{8\pi^2}} g_3^2(\mu_0 )\, {\rm ln }(\mu /\mu_0 )}}\ .
\eea

\vspace{.3truecm}
\noindent
We have solved the equations for the top Yukawa coupling $h_t$ and
Higgs self-coupling $\lm $ numerically using the experimental values of
$g_1^2=0.13, g_2^2=0.42, g_3^2=1.46$ \cite{PDG} and $h_t=1.01$
(for $m_t=176$~GeV from CDF),
$h_t=1.14$ (for $m_t=199$~GeV from D0) \cite{CDF} at the electroweak scale.
After symmetry breaking the masses of the particles are related
to the couplings as
$m_W^2={\frac 12}g_2^2v^2$, $m_Z^2={\frac 12}(g_1^2+g_2^2)v^2$,
$m_H^2=2\lm v^2$ and $m_t=h_t v$ with $v \simeq 174$ GeV being the
vacuum expectation value of the Higgs field.

Taking Eqs.(2)-(4) into account we have found the running mass of the
Higgs as a function of the large scale $\Lambda$, where the formula
(\ref{eqvelt}) is assumed to be valid.
The results for the top masses 176 GeV and 199 GeV are given
in Fig.~1.
It is seen that for the scale of new physics being at
$\Lambda =10^{15} - 10^{19}$~GeV and with
$m_t=176$~GeV (199~GeV) at the electroweak scale, one obtains for the Higgs
mass $m_H \sim 170$~GeV (210~GeV). With lower than $10^{15}-10^{19}$~GeV
values for $\Lambda$, however, the Higgs mass $m_H$ increases. If in
addition one imposes that the quadratic corrections to the
Higgs mass, i.e.

\bea
\Delta m_H^2 = \left( \frac 32 g_1^2 +\frac 92 g_2^2
+ 6 \lambda - 12 h_t^2 \right) \frac {\Lambda^2}{16 \pi^2}\ ,
\eea
can become at most equal to the {\it physical} mass value \cite{mV}
(``naturalness"), then for $m_t=176$ GeV (199 GeV)
one  obtains $\Lambda \simeq 1.8$ TeV and, correspondingly,
a Higgs mass around 260 GeV (300 GeV) at the electroweak scale (see Fig.~1).
Notice also that if one would impose the
 relation (1) to be valid at the electroweak scale one would obtain higher
values for the Higgs mass, namely $m_H=320$~GeV (370~GeV).

Thus we see that under the assumption that quadratic divergences in the
standard model are cancelled at GUT-Planck mass scale the Higgs mass
is expected to be in the range 170~-~210~GeV for $m_t \sim 176-199$~GeV.
If one insists however on the naturalness of the theory one obtains
an upper bound on the scale, $\Lambda\ \lsim \ 1.8$~TeV, and
correspondingly, a lower bound on the Higgs mass $m_H\ \gsim \ 260$~GeV
for $m_t \ \gsim \ 176$~GeV.

\section{Coleman-Weinberg mechanism for spontaneous symmetry breaking}

As first pointed out by Coleman and Weinberg \cite{CW} the spontaneous
symmetry breaking may be driven by radiative corrections in theories
which at the tree level do not exhibit such breaking. The advantage of
this dynamical mechanism is that the symmetry breaking does not have
to be put in by hand. In the framework
of the latter mechanism it has been recently argued \cite{FGHW} that
top quark loops
may trigger the symmetry breaking in the standard electroweak model and
as a consequence, the Higgs boson mass is expected
to be $m_H \leq 400$ GeV depending on the value of the top quark mass
and the physical cutoff $\Lambda$.

Here we shall study the implications of  the
Coleman-Weinberg mechanism for the spontaneous symmetry breaking
in combination with
the cancellation of quadratic divergences in the standard
model. The starting point is the one-loop effective potential which
includes the one-loop top quark and gauge boson contributions. The
latter is easily calculated and the result is well-known \cite{CW,FGHW}.
We have

\bea
V &=& m^2 \phi^2 + \frac{1}{32\pi^2}\int_{0}^{\Lambda^2} dq^2
q^2 \left\{ 6\ \ln \left( 1+ \frac{g_2^2 \phi^2}{2q^2}\right) +
3\ \ln \left( 1+ \frac{(g_1^2+g_2^2) \phi^2}{2q^2}\right) \right.
\nonumber \\
  & & \left. - 12\ \ln \left( 1+ \frac{h_t^2 \phi^2}{q^2}\right) \right\} .
\eea

Note that we have not included
quartic scalar self-interactions, i.e. we start with a simple Lagrangian
of a massive scalar $\phi$ ($m^2 > 0$) interacting with a massless
fermion (top quark). Thus the scalar self-interactions will be induced
by quantum corrections. We postpone comments on this point to the end of
this section.

After performing the integrals in (6) and neglecting terms that vanish as
$\Lambda \rightarrow \infty $ we obtain finally

\bea
V &=& m^2 \phi^2 + \left( \frac{3}{2}g_1^2
+ \frac{9}{2}g_2^2 -12h_t^2 \right) \frac{\Lambda^2\phi^2}{32\pi^2}
+ \frac{1}{64\pi^2} \left\{\frac{3}{2}g_2^4\phi^4
\left( \ln\frac{g_2^2\phi^2}{2\Lambda^2} - \frac{1}{2}\right)
\right. \nonumber \\
  & & \left. + \frac{3}{4}(g_1^2+g_2^2)^2\phi^4
\left( \ln\frac{(g_1^2+g_2^2)\phi^2}{2\Lambda^2} - \frac{1}{2}\right)
 - 12h_t^4\phi^4
\left( \ln\frac{h_t^2\phi^2}{\Lambda^2} - \frac{1}{2}\right)\right\} \ .
\eea

To extend the region of validity of the one-loop effective potential we
can use its RG improved version \cite{Sher,FJS} and with this aim
 we shall run the couplings in Eq.(7) according to  their RGEs
given by Eqs.(2)-(4).

Next we impose that the quadratic divergences (proportional to
$\Lambda^2$) in
the RG improved effective potential (7) are cancelled. This implies
the relation

\be
\frac 32 g_1^2 +\frac 92 g_2^2  - 12 h_t^2 = 0 \ .
\ee

Since the latter equation is scale-dependent, we should require it to
be valid at some fixed scale. Notice that if one
would assume relation (8) to be valid at the electroweak scale one
would obtain a light top quark ($m_t^2=(m_Z^2+2 m_W^2)/4$, i.e.
$m_t \simeq 75$~GeV), which is experimentally excluded. As in our
previous analysis, a natural choice for the scale will be
 the cutoff scale $\Lambda$ where new physics enters into play.

At low energies the effective potential (7) develops a new minimum
$\langle \phi \rangle = v \neq 0$ and thus the symmetry is spontaneously
broken due to quantum corrections. The Higgs mass at the one-loop level
will be given by

\bea
m_H^2 =\left. \frac{1}{2}\frac{\partial^2 V}{\partial \phi^2}\right|_{
\phi=v} &=& \frac{3}{32 \pi^2 v^2}\left\{ 4m_t^4\left(
2\ \ln\frac{\Lambda^2}{m_t^2} - 1\right) - m_Z^4\left(
2\ \ln\frac{\Lambda^2}{m_Z^2} - 1\right)  \right.\nonumber \\
        & & \left. - 2 m_W^4\left( 2\ \ln\frac{\Lambda^2}{m_W^2}
- 1\right)\right\} \ .
\eea

The numerical procedure for evaluating the top quark and the Higgs mass
at the electroweak scale is then as follows: we require Eq.(8) to be
valid at some scale $\Lambda$ and solve the RGE for the top Yukawa
coupling (given in Eqs.(2)-(4)) to find the top mass at the electroweak
scale. Finally, we evaluate the Higgs mass by using Eq.(9). The results
for $m_t$ and $m_H$ as a function of the scale are given in Fig.~2
assuming $\alpha_s(m_Z)=0.116$ \cite{PDG}. We
notice that in order to have a top quark mass in agreement with the recent
experimental results \cite{CDF} the scale $\Lambda$ at which relation (8)
is valid should be sufficiently high and at most of the order of the Planck
scale, i.e. $\Lambda \sim 10^{19}$~GeV. In the latter case the Higgs mass
will be $m_H\ \lsim \ 300$~GeV, while for the top quark mass we obtain
$m_t\ \lsim \ 150$~GeV.
The above results are of course sensitive to the initial value of the strong
coupling constant $\alpha_s$. For instance, if we use the value
$\alpha_s(m_Z)=0.123 \pm 0.006$ from LEP event shapes \cite{PBR}, we obtain
the upper bounds $m_H\ \lsim \ 330$~GeV and $m_t\ \lsim \ 155$~GeV.

It is worth mentioning that we have used in our calculations the running
masses for the Higgs scalar and top quark. The physical pole masses can
be computed from the running ones through the corresponding corrections
which are typically of the order of a few percent. In the case of the
top quark the latter corrections increase the value of $m_t$ by about 7~\%,
thus implying $m_t\ \lsim \ 160-170$~GeV\renewcommand{\thefootnote}{*}
\footnote{It is interesting that a preliminary top mass result from the
dilepton channels reported by D0 collaboration is $m_t=145 \pm 25 \pm 20$~GeV
\cite{BK}.} depending on the value of $\alpha_s$.

It is obvious that in the Coleman-Weinberg case the naturalness
(i.e. the requirement that the quadratic corrections to the mass are
smaller than the physical Higgs mass in (9)) cannot be required since this
would lead to a cutoff scale in the TeV range and thus $m_t$ would be too
low (about 90 GeV, cf. Fig.~2), which is excluded by the recent
experimental limits on the top quark mass.

Finally, let us comment on the quartic scalar self-coupling $\lambda$. We
have assumed it to be zero at $\Lambda$ scale (cf. Eq.(8)). At the
electroweak scale it is defined as

\be
\lambda=\frac{1}{12}\left. \frac{\partial^4 V}{\partial\phi^4}
\right|_{\phi=v} \ ,
\ee
where $V$ is the effective potential given in Eq.(7). (We have chosen
the coefficient in the definition of $\lambda$ so that the quartic term
in the effective potential is $\lambda \phi^4 /2$).
 Then using Eqs.(9) and (10) it is straightforward to show that in the
leading ln~$\Lambda$ approximation $m_H^2 = 2 \lambda v^2$, which is
the same
expression as in the usual mechanism for spontaneous symmetry breaking.

To conclude, our suggestion in this letter is that if the standard model
could be rendered free of quadratic divergences, then the cancellation
of such divergences should occur
at the scale of new physics and not at the electroweak scale.
As a consequence  of this approach the mass relations between $m_H$
and $m_t$ are drastically changed at the electroweak scale in the
direction of lowering the Higgs mass.
\vspace*{1 cm}

We are grateful to Claus Montonen and Risto Orava for useful discussions.

\input{epsf.sty}
\begin{figure}[t]
\leavevmode
\begin{center}
\mbox{\epsfxsize=15.cm\epsfysize=15.cm\epsffile{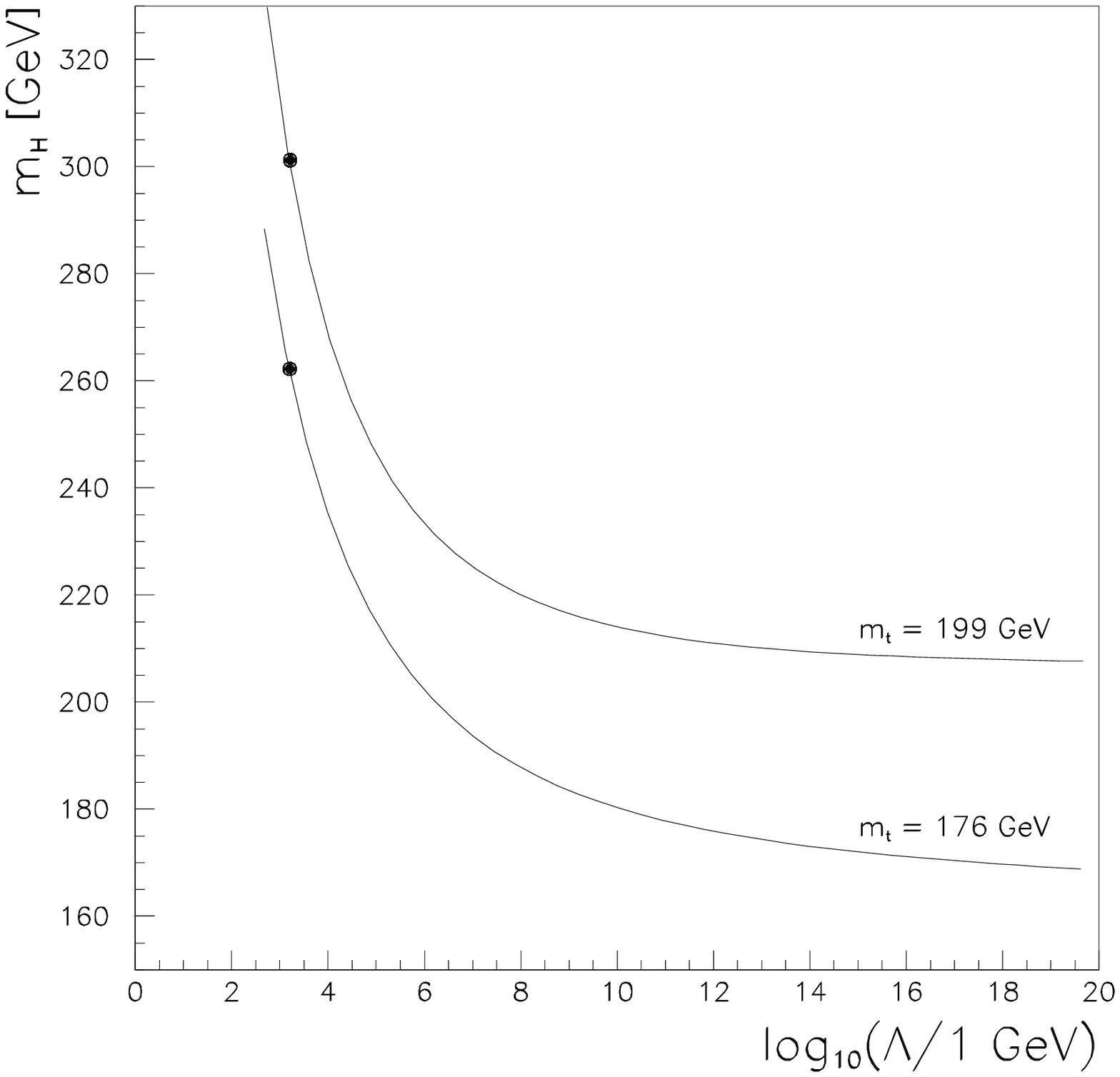}}
\end{center}
\caption{
Higgs mass $m_H$ as a function of the scale $\Lambda$ where cancellation
of quadratic divergences is assumed.
The bullets denote the intersection points at which the quadratic
corrections $\Delta m_H $ (cf. Eq.(5)) equal the physical mass $m_H$.
}
\label{fig1}
\end{figure}

\begin{figure}[t]
\leavevmode
\begin{center}
\mbox{\epsfxsize=15.cm\epsfysize=15.cm\epsffile{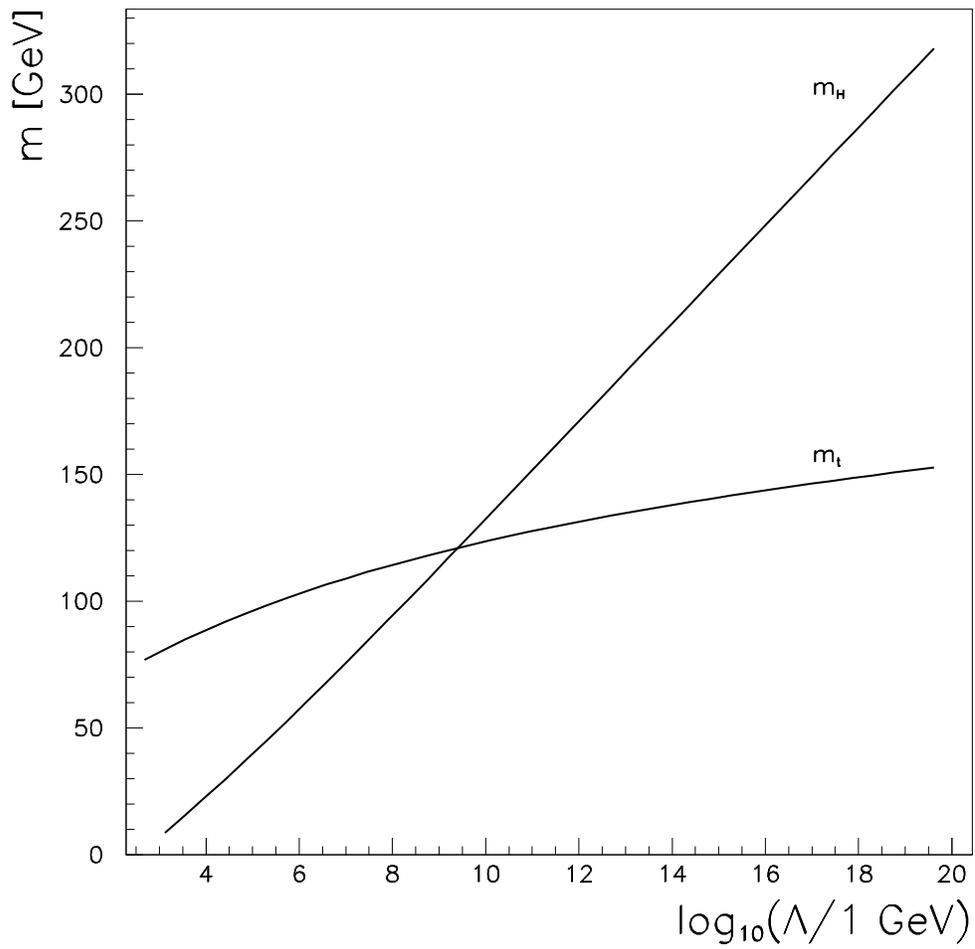}}
\end{center}
\caption{
Higgs mass $m_H$ and top quark mass $m_t$ as functions of the scale
$\Lambda$ in the case of spontaneous symmetry breaking via Coleman-Weinberg
mechanism.
}
\label{fig2}
\end{figure}

\end{document}